\documentclass[fleqn,10pt]{wlscirep} 
\usepackage{algorithm}
\usepackage[noend]{algpseudocode}
\usepackage{lineno}

\title{Efficient collective influence maximization in cascading processes with first-order transitions}
\author[1]{Sen Pei}
\author[2]{Xian Teng}
\author[1]{Jeffrey Shaman}
\author[2]{Flaviano Morone}
\author[2,*]{Hern{\'a}n A. Makse}
\affil[1]{Department of Environmental Health Sciences, Mailman School of Public Health, Columbia University, New York, NY 10032, USA}
\affil[2]{Levich Institute and Physics Department, City College of New York, New York, NY 10031, USA}

\affil[*]{hmakse@lev.ccny.cuny.edu}


\begin{abstract}
In many social and biological networks, the collective dynamics of the entire system can be shaped by a small set of influential units through a global cascading process, manifested by an abrupt first-order transition in dynamical behaviors. Despite its importance in applications, efficient identification of multiple influential spreaders in cascading processes still remains a challenging task for large-scale networks. Here we address this issue by exploring the collective influence in general threshold models of cascading process. Our analysis reveals that the importance of spreaders is fixed by the subcritical paths along which cascades propagate: the number of subcritical paths attached to each spreader determines its contribution to global cascades. The concept of subcritical path allows us to introduce a scalable algorithm for massively large-scale networks. Results in both synthetic random graphs and real networks show that the proposed method can achieve larger collective influence given the same number of seeds compared with other scalable heuristic approaches.
\end{abstract}
\begin{document}

\flushbottom
\maketitle
\thispagestyle{empty}

\section*{Introduction}

Cascading process lies at the heart of an array of complex phenomena in social and biological systems, including failure propagation in infrastructure \cite{Buldyrev2010}, adoption of new behaviors \cite{Watts2007a}, diffusion of innovations in social networks \cite{Rogers1995} and cascading failures in brain networks \cite{Reis2014}, etc. In these cascading processes, a small number of influential units, or influencers, arise as a consequence of the structural diversity of the underlying interacting networks \cite{Kleinberg2007}. In different fields, it has been accepted that the initial activation of a small set of such ``superspreaders'', who usually hold prominent locations in networks, is capable of shaping the collective dynamics of large populations \cite{Domingos2001,Valente1999,Galeotti2009}. In practice, identification of superspreaders can help to control the entire network's dynamics with a low cost, e.g., a company can boost product popularity by targeted advertisement on influencers in viral marketing, or we can maintain the robustness of infrastructure systems by protecting structurally pivotal units. Given its great practical values in a wide range of important applications, the problem of locating superspreaders has attracted much attention in various disciplines \cite{Kempe2003,Leskovec2007,Chen2009,Kitsak2010,Morone2015,Morone2016,Pei2013,Pei2014,Altarelli2013a,Altarelli2013,Guggiola2015,Mugisha2016,Braunstein2016,Teng2016,Pei2015,Radicchi2016,Hu2015,Lawyer2015,Quax2013,Tang2015}.

In the simple case of finding single influential spreaders, centrality-based heuristic measures such as degree \cite{Albert2000}, Betweenness \cite{Freeman1979}, PageRank \cite{Brin1998} and K-core \cite{Seidman1983} are routinely adopted. Beyond this non-interacting problem of finding single spreaders, it becomes more complicated when trying to select a group of spreaders, due to the collective effects of multiple agents. In fact, searching for the optimal set of influencers in cascading dynamics is an NP-hard problem and remains to be a challenging conundrum in network science \cite{Kempe2003}. To address the many-body problem, several approaches designed for the influence maximization in different models have been proposed. In the case of percolation model, a framework for optimal percolation based on the stability analysis of zero solution was developed \cite{Morone2015,Morone2016}. More recently, message passing algorithms for optimal decycling in statistical physics further pushed the critical point toward its optimal value \cite{Mugisha2016,Braunstein2016}. For susceptible-infected-recovered (SIR) model, the problem of finding influential spreaders is also explored using the percolation theory in recent works \cite{Radicchi2016,Hu2015}. In the above models, cascading processes can be transformed to the percolation model with a continuous phase transition. While optimal percolation theory \cite{Morone2015,Morone2016} applies only to systems with second order phase transitions, here we treat the case of cascading models which present first order discontinuous transitions. Such transitions cannot be treated with the stability analysis methods based on the non-backtracking matrix as done in \cite{Morone2015,Morone2016} for models with continuous transitions, so a new approach is needed.

In a large variety of contexts, the cascading process is properly described by the Linear Threshold Model (LTM) in which the states of nodes are determined by a threshold rule \cite{Granovetter1978,Schelling1978,Valente1995,Watts2002,Baxter2010,Goltsev2006,Dorogovtsev2006,Schwarz2006}. That is, a node will become active only after a certain number of its neighbors have been activated. The choice of threshold $m_i = k_i-1$ in LTM guarantees a continuous phase transition, where $k_i$ is the degree of node $i$. In this case, the cascading dynamics of LTM can be mapped to the classical percolation process \cite{Morone2015}, for which the influence maximization problem can be solved by various algorithms designed for optimal percolation. Nevertheless, for other choices of threshold, LTM exhibits a first-order (i.e., discontinuous) phase transition. In fact, influence maximization in LTM corresponds to finding nontypical trajectories of cascading processes that deviate from the average ones \cite{Altarelli2013a}. Altarelli {\it et al.} \cite{Altarelli2013a} analyzed the statistics of large deviations of LTM dynamics with a belief propagation algorithm, and further developed a Max-Sum (MS) algorithm to explicitly find the optimal set of seeds in terms of a predefined energy function \cite{Altarelli2013}. Guggiola and Semerjian \cite{Guggiola2015} obtained the theoretical limit of the size of minimal contagious sets for random regular graphs, and used a survey propagation like algorithm to locate the minimal set of seeds. Given these recent progresses in searching for optimal influencers in LTM, it is a challenging task to apply these methods to massively large-scale networks with tens of millions nodes encountered in modern big-data analysis. Therefore, the problem of developing an efficient scalable algorithm of influence maximization in cascading models with discontinuous transitions that is feasible in real-world applications still needs to be further explored.

Here, we examine the collective influence in general LTM, and develop a scalable algorithm for influence maximization. By analyzing the message passing equations of LTM, we formulate the form of interactions between spreaders and provide an analytical expression of their contributions to cascading process. Each seed's contribution, defined as the {\it collective influence in threshold model} (CI-TM), is determined by the number of subcritical paths emanating from it. Since the subcritical paths are such routes along which cascades can propagate, CI-TM can be considered as a reliable estimation of seeds' structural importance in LTM. CI-TM is the generalization of the CI algorithm of optimal percolation for second order transitions treated in \cite{Morone2015,Morone2016} to the present case of first order transitions. To apply our method to large-scale networks, we present an efficient adaptive selection procedure to achieve collective influence maximization. Compared with other competing heuristics, our results on both synthetic and realistic large-scale networks reveal that the proposed mechanism-based algorithm can produce a larger cascading process given the same number of seeds. 

\section*{Results}
\subsection*{Collective influence in threshold models: CI-TM}

\begin{figure}
\centering
\includegraphics[width=0.8\linewidth]{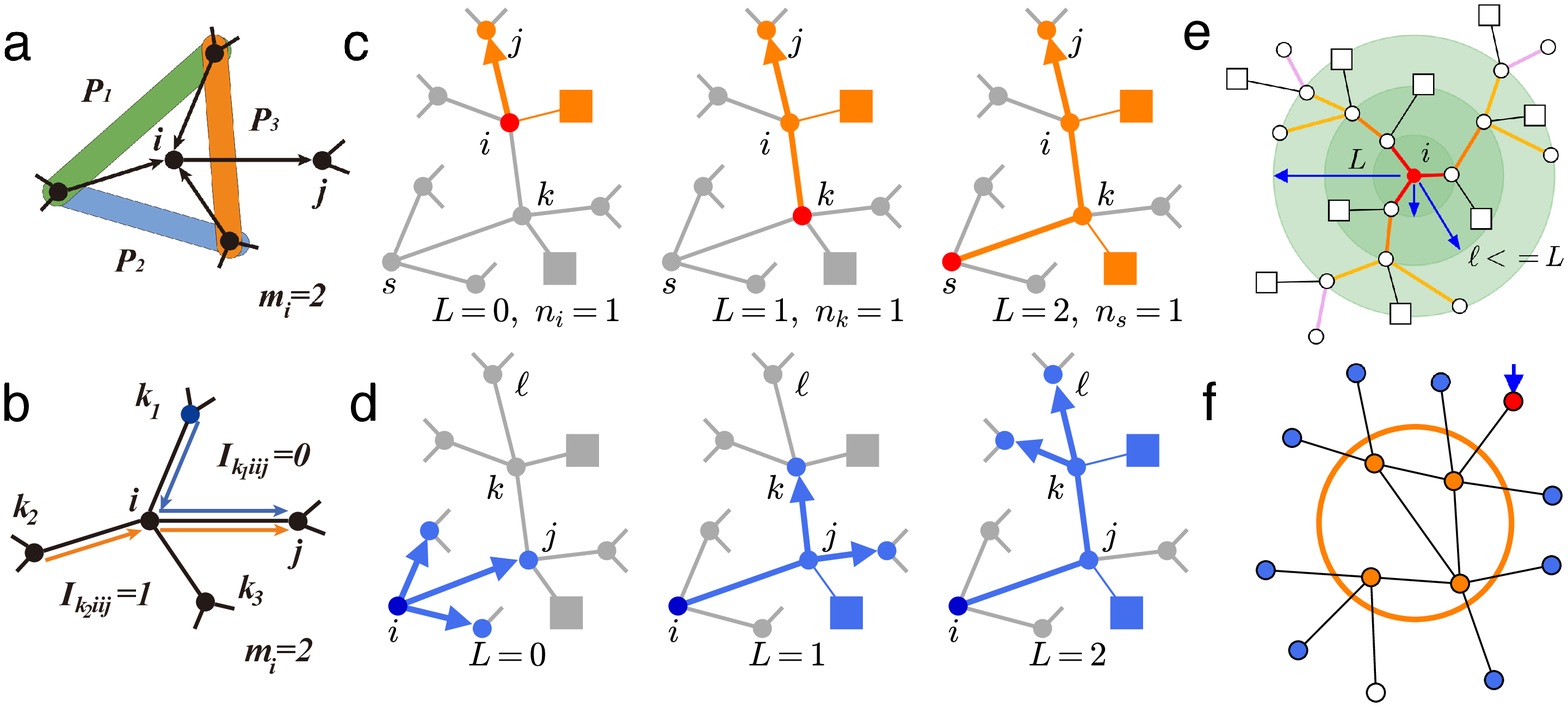}\\
\caption{{\bf Subcritical paths and collective influence of spreaders.} (a), Three combinations of neighbors $P_1$, $P_2$ and $P_3$ corresponding to $\nu_{i\to j}$ in message passing equation. Node $i$ has a threshold $m_i=2$. The full activation of at least one combination will lead to $\nu_{i\to j}=1$.(b), For link $i\to j$ with an active neighbor $k_1$ and inactive ones $k_2$ and $k_3$, $I_{k_1\to i, i\to j}=0$ since $i$ has $0$ ($<m_i-1$) active neighbor excluding $k_1$ and $j$, while $I_{k_2\to i, i\to j}=1$ because $i$ has $1$ ($=m_i-1$) active neighbor $k_1$ excluding $k_2$ and $j$. (c), Illustrations of subcritical paths ending with link $i\to j$ for $L=0,\ 1,\ 2$. Red dots stand for seeds, while squares represent $m-1$ active neighbors attached to subcritical nodes. Subcritical paths are highlighted by thick links. (d), The contribution of seed $i$ to $\|\nu_{\to}\|$ exerted through subcritical paths of length $L=0,\ 1,\ 2$. (e), Calculation method of $\text{CI-TM}_L(i)$. Subcritical paths starting from $i$ with length $\ell\leq L$ are displayed by different colors. (f), An example of subcritical cluster. Assuming a uniform threshold $m=3$, nodes inside the circle are subcritical since they all have $2$ active neighbors, represented by blue nodes. Activation of the red node will trigger a cascade covering all subcritical nodes.}\label{illustration}
\end{figure}

We present a theoretical framework to analyze the collective influence of individuals in general LTM. For a network with $N$ nodes and $M$ links, the topology is represented by the adjacency matrix $\{A_{ij}\}_{N\times N}$, where $A_{ij}=1$ if $i$ and $j$ are connected, and $A_{ij}=0$ otherwise. The vector $\mathbf{n}=(n_1, n_2, \cdots, n_N)$ records whether a node $i$ is chosen as a seed ($n_i=1$) or not ($n_i=0$). The total fraction of seeds is therefore $q = \sum_i n_i/N$. During the spreading, the state of each node falls into the category of either active or inactive. The spreading starts from a $q$ fraction of active seeds and evolves following a threshold rule: a node $i$ becomes active when $m_i$ neighbors get activated. This process terminates when there are no more newly activated nodes. We introduce $\nu_i$ as node $i$'s indicator in active ($\nu_i=1$) or inactive ($\nu_i=0$) state at the final stage, and denote $Q(q)$ as the size of the giant connected component of active population.

For a directed link $i\to j$, we introduce $\nu_{i\to j}$ as the indicator of $i$ being in an active state assuming node $j$ is disconnected from the network. If $n_i=1$, then $\nu_{i\to j}=1$. Otherwise, $\nu_{i\to j}=1$ only when there are at least $m_i$ active neighbors excluding $j$. Since there exist many possible choices of these $m_i$ neighbors, we define $P_{\partial i\setminus j}^{m_i}$ as the set of all combinations of $m_i$ nodes selected from $\partial i\setminus j$, where $\partial i\setminus j$ is the set of nearest neighbors of $i$ excluding $j$. Clearly, if $i$ has $k_i$ connections emanating from it, there are $\binom{k_i-1}{m_i}$ combinations, so the set $P_{\partial i\setminus j}^{m_i}$ contains $\binom{k_i-1}{m_i}$ elements, denoted by $P_h$, $h=1, \cdots, \binom{k_i-1}{m_i}$. Each element $P_h$ has the form  $P_h=\{p_{h_1}, \cdots, p_{h_{m_i}}\}$ where $\{p_{h_1}, \cdots, p_{h_{m_i}}\}$ are the $m_i$ nodes in the $h$th combination. Figure \ref{illustration}a illustrates all three combinations $P_1, P_2$ and $P_3$ corresponding to $\nu_{i\to j}$ for node $i$ with a threshold $m_i=2$. Should at least one combination is fully activated, we have $\nu_{i\to j}=1$.

Generally, for locally tree-like networks, we have the following message passing equation:
\begin{equation}\label{mpvij}
\nu_{i\to j}=n_i+(1-n_i)[1-\prod_{P_h\in P_{\partial i\setminus j}^{m_i}}(1-\prod_{p\in P_h}\nu_{p\to i})].
\end{equation}
The final state of $i$ is given by
\begin{equation}\label{mpvi}
\nu_{i}=n_i+(1-n_i)[1-\prod_{P_h\in P_{\partial i}^{m_i}}(1-\prod_{p\in P_h}\nu_{p\to i})].
\end{equation}
The above equations Eq. (\ref{mpvij}-\ref{mpvi}) describe the general cases of LTM. For the special choice of threshold $m_i = k_i-1$, there is only one combination in $P_{\partial i\setminus j}^{m_i}$, and the transition becomes continuous, and then it can be treated with the stability methods of optimal percolation as done in \cite{Morone2015,Morone2016}. We note that, Eq. (\ref{mpvij}-\ref{mpvi}) are only valid under the locally tree-like assumption. For synthetic random networks, this assumption holds since short loops appear with a probability of order $O(1/N)$ \cite{Morone2015}. Nevertheless, a considerable number of short loops may exist in real-world networks. For those networks with clustering, many prior works have confirmed that results obtained for tree-like networks apply quite well also for loopy graphs. For instance, Melnik {\it et al.} found that, for a series of problems, the tree-like approximation worked well for clustered networks as long as the mean intervertex distance was sufficiently small \cite{Melnik2011}. As most of real-world networks are small-world, the approximation of Eq. (\ref{mpvij}-\ref{mpvi}) should be reasonable provided the density of loops is not excessively large.

For all the $2M$ directed links $i\to j$,  Eq. (\ref{mpvij}) is a nonlinear function of $\nu_\to=(\cdots,\nu_{i\to j}, \cdots)^T$:
\begin{equation}\label{fullmp}
\nu_{\to}=\mathbf{n_{\to}}+\mathbf{G}(\nu_\to).
\end{equation}
In Eq. (\ref{fullmp}), $\mathbf{n_\to}=(\cdots, n_{i\to j}, \cdots)^T$ in which $n_{i\to j}=n_i$ for link $i\to j$, and $\mathbf{G}=(\cdots, G_{i\to j}, \cdots)^T$ where $G_{i\to j}$ is the nonlinear function of vector $\nu_\to$ for link $i\to j$. Given the initial configuration of seeds $\mathbf{n}$, the final state of $\nu_{\to}$ is fully determined by the self-consistent Eq. (\ref{fullmp}). Unfortunately, it cannot be solved directly due to the exponentially growing number of combinations in $P_{\partial i\setminus j}^{m_i}$. Therefore, for a small number of seeds, we adopt the iterative method to estimate the solution. In this point of view, Eq. (\ref{fullmp}) can be treated as a discrete dynamical system
\begin{equation}\label{fullupdate}
\nu_{\to}^{t+1}=\mathbf{n_{\to}}+\mathbf{G}(\nu_\to^t)
\end{equation}
with the initial condition $\nu_\to^0=\mathbf{n_{\to}}$.

To simplify the calculation, we approximate the nonlinear function $G_{i\to j}$ by linearization. Define $G'_{i\to j}(\nu_\to)=(\cdots,\frac{\partial G_{i\to j}}{\partial\nu_{k\to\ell}}, \cdots)$. By Eq. (\ref{mpvij}), we know that $\frac{\partial G_{i\to j}}{\partial\nu_{k\to\ell}}=0$ for $\ell\neq i$. While in the case of $\ell=i$ and $k\neq j$, we have
\begin{equation}\label{calculateF}
\frac{\partial G_{i\to j}}{\partial\nu_{k\to i}}=(1-n_i)\prod_{\bar{P}_h\in P_{\partial i\setminus j}^{m_i}, k\not\in \bar{P}_h}(1-\prod_{p\in \bar{P}_h}\nu_{p\to i})\times\sum_{P_h\in P_{\partial i\setminus j}^{m_i}, k\in P_h}[(\prod_{p\in P_h\setminus k}\nu_{p\to i})\prod_{P'_h\neq P_h, k\in P'_h}(1-\prod_{p\in P'_h}\nu_{p\to i})].
\end{equation}

Although Eq. (\ref{calculateF}) has a complex form, in fact it is only determined by a simple quantity $a_{k\to i, i\to j}=\sum_{p\in\partial i\setminus (k, j)}\nu_{p\to i}$, which is interpreted as the number of $i$'s active neighbors excluding $k$ and $j$ when $i$ is absent from the network. On one hand, if $a_{k\to i, i\to j}\geq m_i$, at least one term of $\prod_{p\in \bar{P}_h}\nu_{p\to i}$ equals one, since we are selecting $m_i$ elements from a set containing at least $m_i$ elements of value 1. Under such condition, $\frac{\partial G_{i\to j}}{\partial\nu_{k\to i}}=0$. On the other hand, if $a_{k\to i, i\to j}\leq m_i-2$, all the terms $\prod_{p\in P_h\setminus k}\nu_{p\to i}$ are zeros because we are selecting $m_i-1$ elements from a set containing at most $m_i-2$ nonzero elements, which also leads to $\frac{\partial G_{i\to j}}{\partial\nu_{k\to i}}=0$. When $a_{k\to i, i\to j}=m_i-1$, all the terms $\prod_{p\in \bar{P}_h}\nu_{p\to i}$ and $\prod_{p\in P'_h}\nu_{p\to i}$ are zeros, and only the exact combination of these $m_i-1$ nonzero elements would lead to $\prod_{p\in P_h\setminus k}\nu_{p\to i}=1$. Therefore, we have $\frac{\partial G_{i\to j}}{\partial\nu_{k\to i}}=1-n_i$. Based on the above reasoning, we define a quantity $I_{k\to \ell, i\to j}$ for links $k\to \ell$ and $i\to j$ as follows:
\begin{equation}\label{I}
I_{k\to \ell, i\to j}=
\begin{cases}
1 &\text{if } \ell=i, k\neq j, a_{k\to i, i\to j}=m_i-1,\\
0 &\text{otherwise}.
\end{cases}
\end{equation}

The definition of $I_{k\to \ell, i\to j}$ is reminiscent of the Hashimoto non-backtracking (NB) matrix $\mathcal{B}$ \cite{Morone2015,Hashimoto1989,Martin2014}. In the case of $m_i=k_i-1$, our quantity $I_{k\to \ell, i\to j}$ can be transformed to the corresponding element of NB matrix $\mathcal{B}_{k\to \ell, i\to j}$. In fact, $I_{k\to \ell, i\to j}$ is closely related to the concept of subcritical nodes. Recall that a node $i$ is subcritical if it has $m_i-1$ active neighbors \cite{Goltsev2006,Dorogovtsev2006,Schwarz2006}. This implies that one more activation of its neighbors will cause $i$ activated. From Eq. (\ref{I}) we know that $I_{k\to \ell, i\to j}=1$ only if the links $k\to \ell$ and $i\to j$ are connected, non-backtracking, and additionally, node $i$ is subcritical in the absence of node $k$ and $j$. In Fig. \ref{illustration}b, node $i$ has an active neighbor $k_1$ and two inactive ones $k_2$ and $k_3$. By definition, for a threshold $m_i=2$, we conclude $I_{k_1\to i, i\to j}=0$ since $i$ has no active neighbor excluding $k_1$ and $j$, while $I_{k_2\to i, i\to j}=1$ because $i$ has 1 $(=m_i-1)$ active neighbor excluding $k_2$ and $j$.

For a small $\nu_{\to}$, a standard linearization around origin $\mathbf{0}$ gives $G_{i\to j}(\nu_\to)\approx G_{i\to j}(\mathbf{0})+G'_{i\to j}(\mathbf{0})\nu_\to$. However this will cause degeneracy since Eq. (\ref{calculateF}) constantly gives $G'_{i\to j}(\mathbf{0})=\mathbf{0}$. Therefore, we approximate $G_{i\to j}(\nu_\to)$ by $G_{i\to j}(\mathbf{0})+G'_{i\to j}(\nu_\to)\nu_\to$ given $\nu_\to$ is close to $\mathbf{0}$. In Methods, we prove that this linearization has an approximation accuracy of $O(|\nu_{\to}|^2)$ ($|\cdot|$ is the vector norm), same as the linear Taylor expansion. To account for the increasing network size as $N\to\infty$, we define the vector norm as $|\nu_{\to}|\equiv\sum_{ij}\nu_{i\to j}/2M$ ($2M$ is the number of directed links), so that $|\nu_{\to}|$ is always bounded below 1 for all network sizes. The linear approximation is valid when the number of links attached to initial seeds is small compared with all directed links. As we will see, the fraction of seeds at the discontinuous transition is small for both synthetic and realistic networks. Therefore, the linear approximation before the critical point should be valid. In Methods, we compare $|\nu_{\to}|$ calculated by linear approximation with its real value on a scale-free network assuming LTM is initialed by one single seed. The approximation results agree well with their true values. See the detailed discussion in Methods.

Combining all direct links, Eq. (\ref{fullupdate}) can be approximated by a linear equation
\begin{equation}\label{update}
\nu_{\to}^{t+1}=\mathbf{n_{\to}}+\mathcal{F}^t\nu_\to^t,
\end{equation}
where $\mathcal{F}^t=(\cdots, G'_{i\to j}({\nu_\to^t}), \cdots)^T$ is a $2M\times2M$ matrix defined on the directed links $k\to\ell$, $i\to j$ with elements
\begin{equation}\label{Felement}
\mathcal{F}^t_{k\to\ell, i\to j}=\left.\frac{\partial G_{i\to j}}{\partial\nu_{k\to\ell}}\right|_{\nu_{\to}^t}.
\end{equation}
With the notion of $I_{k\to \ell, i\to j}$, we can write $\mathcal{F}^t$ as:
\begin{equation}\label{Fdefinition}
\mathcal{F}^t_{k\to\ell, i\to j}=(1-n_i)I^t_{k\to \ell, i\to j}.
\end{equation}

Now we update the state of $\nu_\to^t$ following Eq. (\ref{update}). In the following calculation, we simplify $\mathcal{F}^t_{k\to\ell, i\to j}$ and $I^t_{k\to \ell, i\to j}$ to $\mathcal{F}^t_{k\ell ij}$ and $I^t_{k\ell ij}$ respectively for notation convenience. We put the matrix $\mathcal{F}^t$ in a higher-dimensional space \cite{Morone2015}:
\begin{equation}\label{Mijkl}
\mathcal{F}^t_{k\ell ij}=(1-n_i)A_{k\ell}A_{ij}\delta_{i\ell}(1-\delta_{jk})I^t_{k\ell ij},
\end{equation}
where function $\delta_{i\ell}$ is $1$ if $i=\ell$, and $0$ otherwise. The indices $k, \ell, i, j$ run from $1$ to $N$. Starting from $\nu_{\to}^0=\mathbf{n}_{\to}$, $\nu_{\to}^{1}=\mathbf{n_{\to}}+\mathcal{F}^0\mathbf{n_{\to}}$ gives
\begin{equation}\label{v1}
\nu^1_{i\to j}=n_i+(1-n_i)A_{ij}\sum_k A_{ki}(1-\delta_{jk})I^0_{kiij}n_k.
\end{equation}
The physical meaning of Eq. (\ref{v1}) can be interpreted as follows. If node $i$ is a seed, $\nu_{i\to j}^1=1$. Otherwise, $\nu_{i\to j}^1$ is nonzero if $i$ is subcritical ($I^0_{kiij}=1$) and at least one of its corresponding neighbors $k$ is a seed ($n_k=1$). Supposing $i$ is not a seed, the contribution of a neighboring seed $k$ is conveyed by the directed path $k\to i \to j$ that satisfies $n_k=1, n_i=0$ and $I^0_{kiij}=1$, which is shown in the second panel of Fig. \ref{illustration}c.

For $t=2$, we have
$\nu_{\to}^{2}=\mathbf{n_{\to}}+\mathcal{F}^1\mathbf{n_{\to}}+\mathcal{F}^1\mathcal{F}^0\mathbf{n_{\to}}$.
Therefore,
\begin{equation}\label{v2}
\nu^2_{i\to j}=n_i+(1-n_i)A_{ij}\sum_k A_{ki}(1-\delta_{jk})I_{kiij}^1n_k+(1-n_i)A_{ij}\sum_{k}(1-n_k)A_{ki}(1-\delta_{jk})I_{kiij}^1\sum_{s}A_{sk}(1-\delta_{is})I_{skki}^0n_s.
\end{equation}
The last term in Eq. (\ref{v2}) is actually the contribution of node $i$'s 2-step neighbors $s$ to $\nu_{i\to j}^2$. The contribution of a seed $s$ is conducted through a directed path $s\to k\to i\to j$ that satisfies $n_s=1, n_k=0, n_i=0$ and $I_{skki}^0=1, I_{kiij}^1=1$ (See Fig. \ref{illustration}c).

Inspired by Eq. (\ref{v2}), we define the concept of {\it subcritical paths}. For a directed link $i\to j$, the path $i_1\to i_2\to \cdots \to i_L\to i\to j$ is a subcritical path of length $L$ if $n_{i_1}=1, n_{i_2}=0, \cdots, n_{i}=0$, $I_{i_1i_2i_2i_3}^0=1, \cdots, I_{i_Liij}^{L-1}=1$, and any two consecutive links are non-backtracking. If $i_1 = i$, we set the path's length $L = 0$. The subcritical paths of length $L=0$, $1$ and $2$ are displayed in Fig. \ref{illustration}c. Notice that, the calculation of $L$-length subcritical paths is in fact implemented by the multiplication of $\mathcal{F}^{L-1}\cdots\mathcal{F}^0$. In fact, the concept of subcritical path has a clear physical meaning. A subcritical path is composed of connected subcritical nodes. So once the node $i_1$ at the beginning of the subcritical path is activated, the cascade of activation will propagate along the path and lead to $\nu_{i\to j}=1$ for the link $i\to j$ at the tail. Therefore, the long-range interaction between node $i_1$ and node $i$ is realized through the subcritical path connecting them. Following this idea, we can generalize Eq. (\ref{v2}) to $\nu_{i\to j}^T$ at a given time $T$. The exact formula for $\nu_{i\to j}^T$ is
\begin{equation}\label{vT}
\nu^T_{i\to j}=n_i+(1-n_i)A_{ij}\sum_{L=1}^{T}\prod_{\ell=1}^{L}\Big[\sum_{k_{\ell}}(1-n_{k_{\ell}}(1-\delta_{\ell L}))A_{k_{\ell}k_{\ell-1}}(1-\delta_{k_{\ell-2}k_{\ell}})I_{k_{\ell}k_{\ell-1}k_{\ell-1}k_{\ell-2}}^{T-\ell}((n_{k_{\ell}}-1)\delta_{\ell L}+1)\Big],
\end{equation}
where $k_{-1}=j$, $k_0=i$ and $k_{\ell}$ runs from $1$ to $N$ for $\ell>0$. Notice that the form of $\nu_{i\to j}^T$ is nothing but $n_i$ plus the contribution of seeds connected to $i$ through subcritical paths with length $L\leq T$ when $n_i=0$.

\subsection*{CI-TM Algorithm}

To quantify the active population in LTM, we define $\|\nu_{\to}\|\equiv 2M|\nu_{\to}|=\sum_{ij}\nu_{i\to j}$, where $2M$ is the total number of directed links. Starting from $\|\nu_{\to}\|=0$ when no seed is selected, $\|\nu_{\to}\|$ increases as more seeds are activated. Therefore, we expect that the collective influence of a given number of seeds can be optimized by maximizing $\|\nu_{\to}\|$.

Based on the form of each element in $\nu_{\to}$, we learn that the contribution of a seed $i$ to $\|\nu_{\to}\|$ is composed of all its collective contributions to every potential element, exerted through the subcritical paths attached to $i$. Therefore, we employ a seed's contribution to $\|\nu_{\to}\|$ to define its Collective Influence in Threshold Model (CI-TM) to find the best influencers. For the trivial case of subcritical paths with length $L=0$, we define $\text{CI-TM}_0(i)=k_i$, where $k_i$ is the degree of node $i$. Thus, at the zero-order approximation we recover the high-degree heuristic. The first panel of Fig. \ref{illustration}d illustrates $\text{CI-TM}_0(i)=3$ for node $i$. For $L\geq1$, subcritical paths are involved in the definition of CI-TM. For $L=1$,
\begin{equation}\label{CI-TM1}
\text{CI-TM}_1(i)=k_i+\sum_{j\in\partial i}(1-n_j)\sum_{k\in\partial j\setminus i} I_{ijjk}^0.
\end{equation}
As shown in Fig. \ref{illustration}(d), the contribution of node $i$ to $\|\nu_{\to}\|$ through subcritical paths of length $L=1$ is 2. Therefore, we have $\text{CI-TM}_1(i)=5$. For $L=2$,
\begin{equation}\label{CI-TM2}
\text{CI-TM}_2(i)=k_i+\sum_{j\in\partial i}(1-n_j)\sum_{k\in\partial j\setminus i}I_{ijjk}^0+\sum_{j\in\partial i}(1-n_j)\sum_{k\in\partial j\setminus i}(1-n_k)I_{ijjk}^0\sum_{\ell\in\partial k\setminus j}I_{jkk\ell}^1
\end{equation}
In Fig. \ref{illustration}d, the additional 2-length subcritical paths also contribute to $\text{CI-TM}_2(i)$, leading to $\text{CI-TM}_2(i)=7$. Moreover, in Fig. \ref{illustration}d, we can observe that for the tree structure, the activation of node $j$ in the first-step update will not affect $I_{jkk\ell}^1$ in the second-step update, which means $I_{jkk\ell}^1=I_{jkk\ell}^0$. More generally, $I_{jkk\ell}$ is not affected by the activation of $k$'s any precedent nodes on the subcritical path. Therefore, we will leave out the superscript $t$ in the definition of CI-TM for locally tree-like networks. We can generalize the above CI-TM calculation to any given $L$. In summary, the definition of node $i$'s influence CI-TM in an area of length $L$ is:
\begin{equation}\label{CI-TM}
\text{CI-TM}_L(i)=\text{number of subcritical paths starting from } i \text{ with length } 0\leq\ell\leq L.
\end{equation}
Figure \ref{illustration}e illustrates the calculation of node $i$'s CI-TM for $L=2$, in which subcritical paths with length $\ell\leq L$ are distinguished by colors. 

For a given fraction $q$ of seeds, our goal is to maximize $\|\nu_{\rightarrow}\|$. As we have explained, the CI-TM value of a seed depends on the choice of other seeds. Therefore, it is hard to obtain the optimal configuration $\{\mathbf{n}|\sum_i n_i/N = q\}$ without turning to extremely time-consuming algorithms. To compromise and obtain a scalable algorithm, we propose an adaptive CI-TM algorithm following a greedy approach. Define $C(i, L)$ as the set of node $i$ plus subcritical vertices belonging to all subcritical paths originating from $i$ with length $\ell\leq L$. Beginning with an empty seed set $S$, we remove the top CI-TM nodes as follow. The calculation proceeds following the CI-TM algorithm.

\begin{algorithm}[H]
\caption{CI-TM algorithm}
\label{CI-TMalgorithm}
\begin{algorithmic}[1]
\State Initialize $S=\emptyset$
\State Calculate $\text{CI-TM}_L$ for all nodes
\For{$l = 1$ to $qN$}
\State Select $i$ with the largest $\text{CI-TM}_L$
\State $S=S\bigcup\{i\}$
\State Remove $C(i,L)$, and decrease the degree and threshold of $C(i,L)$'s existing neighbors by 1
\State Update $\text{CI-TM}_L$ for nodes within $L+1$ steps from $C(i,L)$
\EndFor
\State {\bf end for}
\State Output $S$
\end{algorithmic}
\end{algorithm}

In the above algorithm, we remove $C(i, L)$ once $i$ is added to the seed set. The reason lies in that it is unnecessary to select nodes in $C(i, L)$ as seeds in later calculation, because the activation of $i$ will definitely active $C(i, L)$ (See Fig. \ref{illustration}f). Besides, $C(i, L)$ can be identified during the computation of $\text{CI-TM}_L$ without additional cost. In traditional centrality-based methods, seeds may have significant overlap in their influenced population. It has been reported that the performance of these methods, such as K-core, suffers a lot from this phenomenon \cite{Kitsak2010}. On the contrary, in our algorithm, this problem is alleviated by the removal of subcritical nodes in $C(i, L)$, which successfully reduces the overlap and improves the efficacy of each seed. Although such greedy strategy is not guaranteed to give the exact optimal spreaders, we expect a good performance in comparison with other heuristic methods in large-scale networks. For extreme sparse networks with large numbers of fragmented subcritical clusters, a simple modified algorithm can find a smaller set of influencers (See Methods).

More importantly, the CI-TM algorithm is scalable for large networks with a computational complexity $O(N\log N)$ as $N\to \infty$. On one hand, computing $\text{CI-TM}_L$ is equivalent to iteratively visiting subcritical neighbors of each node layer by layer within $L$ radius. Because of the finite search radius, computing $\text{CI-TM}_L$ for each node takes $O(1)$ time. Initially, we have to calculate $\text{CI-TM}_L$ for all nodes. However, during later adaptive calculation, there is no need to update $\text{CI-TM}_L$ for all nodes. We only have to recalculate for nodes within $L+1$ steps from the removed vertices, which scales as $O(1)$ compared to the network size as $N\to \infty$ as shown in Ref. \cite{Morone2016}. On the other hand, selecting the node with maximal CI-TM can be realized by making use of the data structure of heap that takes $O(\log N)$ time \cite{Morone2016}. Therefore, the overall complexity of ranking $N$ nodes is $O(N\log N)$ even when we remove the top CI-TM nodes one by one. In addition, considering the relative small number of subcritical neighbors, the cost of searching for subcritical paths is far less than that when scanning all neighbors. This permits the efficient computation of CI-TM for considerably large $L$. In our later experiments on finite-size networks, we do not put a limit on $L$ so as to calculate CI-TM thoroughly. But remember that we can always truncate $L$ to speed up CI-TM algorithm for extremely large-scale networks.

\subsection*{Test of CI-TM Algorithm}

\begin{figure}
\centering
\includegraphics[width=0.8\linewidth]{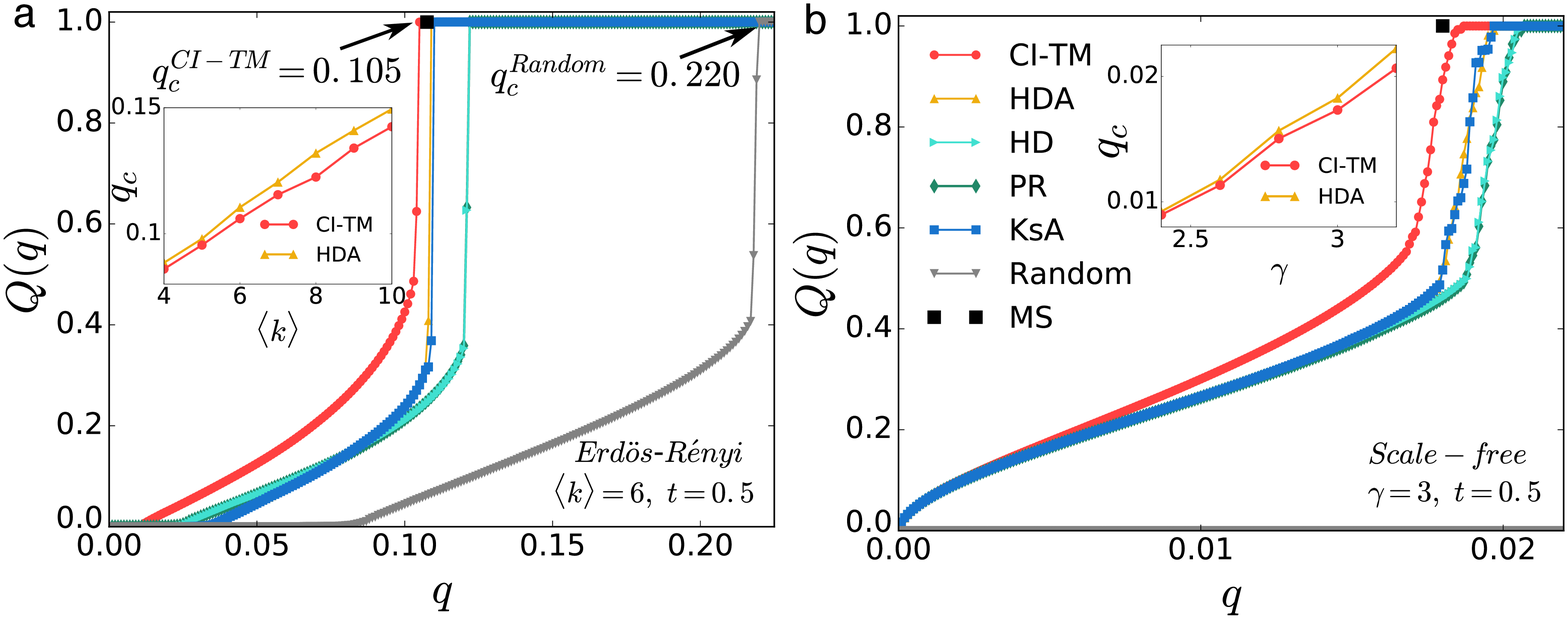}\\
\caption{{\bf Performance of CI-TM algorithm on random networks.} (a), Size of active giant component $Q(q)$ versus the fraction of seeds $q$ for ER networks with size $N=2\times10^5$ and mean degree $\langle k\rangle=6$. Different methods are distinguished by distinct markers and colors. Threshold is set as fractional $t=0.5$. The CI-TM algorithm is run without limitation on $L$. MS is implemented by using $T=40$ and a reinforcement parameter $r=1\times 10^{-4}$. For CI-TM, the identified critical value is $q_c^{\text{CI-TM}}=0.105(1)$ while for Random selection $q_c^{\text{Random}}=0.220(1)$. Inset presents the critical values $q_c$ identified by HDA and CI-TM for different mean degree $\langle k\rangle$. (b), Comparison for scale-free networks with size $N=2\times10^5$, power-law exponent $\gamma=3$, minimal degree $k_{min}=2$ and maximal degree $k_{max}=1000$. The fractional threshold of LTM is also set as $t=0.5$. Inset shows the critical values $q_c$ for different exponents $\gamma$. All the results are averaged over 50 realizations.}\label{random}
\end{figure}

We first simulate LTM dynamics on synthetic random networks, including $\mathrm{Erd\ddot{o}s}$-$\mathrm{R\acute{e}nyi}$ (ER) and scale-free (SF) networks. In the models, we adopt a fractional threshold rule $m_i=\lceil tk_i\rceil$, which means that a node will be activated once $t$ fraction of its neighbors are active ($\lceil\cdot\rceil$ is the ceiling function). Here we choose this special form of threshold setting. But we note that the algorithm can apply to other more general choices of threshold in LTM. In order to verify the efficacy of CI-TM algorithm, we compare its performance against several widely-used ranking methods, including high degree (HD) \cite{Albert2000}, high degree adaptive (HDA), PageRank (PR) \cite{Brin1998} and K-core adaptive (KsA) \cite{Seidman1983}. As a reference, we also display the result of random selection of seeds, as well as the size of optimal seed set identified by Max-Sum algorithm (MS) \cite{Altarelli2013}. Details about these strategies are explained in Methods.

Figure \ref{random}a presents $Q(q)$ versus $q$ on ER networks ($t = 0.5$, $N=2\times10^5$, $\langle k\rangle=6$). Similar to bootstrap percolation on homogeneous networks, $Q(q)$ first undergoes a continuous transition from $Q(q)=0$ to nonzero, and then a first-order transition at a higher value of $q_c$ \cite{Baxter2010}. Remarkably, compared with competing heuristics, CI-TM algorithm achieves a larger active population for a given number of seeds. It not only brings about an earlier continuous transition, but also activates the total population with a smaller seed set. Among all the strategies, random selection represents the average behavior of cascade initiated by randomly chosen seeds, with a critical value $q_c^{\text{Random}}=0.220$. Although the original K-core ranking has an unsatisfactory performance for multi-source spreading \cite{Kitsak2010}, the adaptive version KsA gives a better result similar to HDA. For different threshold $t$, the critical values $q_c$ for CI-TM and other heuristic methods are shown in Table \ref{qcvst}. We also provide the first-order critical value $q_c$ for CI-TM and HDA on ER networks with different average degree $\langle k \rangle$ in the inset of Fig. \ref{random}a. With the growth of $\langle k\rangle$, $q_c$ increases and so does the difference between CI-TM and HDA. In some cases, $q_c$ can be further improved by a simple modification on CI-TM (See Methods).

We then examine CI-TM's performance on SF networks with power-law degree distributions $P(k)\sim k^{-\gamma}$ in Fig. \ref{random}b. We generate SF networks of size $N=2\times10^5$ and power-law exponent $\gamma=3$ following the configuration model \cite{Molloy1995}. It can be seen that the critical value of first-order transition becomes rather small for SF networks, due to the existence of highly connected hubs. Still, CI-TM algorithm outperforms other heuristic approaches by producing a larger active component $Q(q)$ for a given fraction of seed $q$. Since most nodes have a quite small number of connections in SF networks, the cascade triggered by randomly selected seeds is limited to a local scale, even with a relatively large number of seeds, as shown by the grey line at the bottom of Fig. \ref{random}b. This implies, compared with homogeneous networks, the deviations of the optimized trajectories from typical ones are much more extreme in heterogeneous networks. Moreover, as SF networks become more heterogeneous with a smaller power-law exponent $\gamma$ of the degree distribution, the minimal number of seeds required for global cascade decreases accordingly, as shown in the inset of Fig. \ref{random}b.

\begin{table}
\centering
\begin{tabular}{ccccccc}
$t$ & CI-TM & HDA & HD & KsA & PR & Random\\
\hline
0.3 & 0.0197(2) & 0.0258(3) & 0.0266(3) & 0.0260(3) & 0.0264(3) & 0.0566(8) \\
0.4 & 0.0562(2) & 0.0630(4) & 0.0679(5) & 0.0630(3) & 0.0682(4) & 0.1322(8) \\
0.5 & 0.1042(2) & 0.1083(3) & 0.1222(5) & 0.1086(4) & 0.1210(6) & 0.220(1) \\
0.6 & 0.2049(3) & 0.2099(4) & 0.279(1) & 0.2099(5) & 0.282(1) & 0.435(2)
\end{tabular}
\caption{\label{qcvst}
{\bf Critical points for different threshold values.} The critical values $q_c$ found by CI-TM and other heuristics including HDA, HD, KsA, PR, and Random strategies for ER networks ($N=10^5$, $\langle k\rangle=6$) with different threshold values $t$. Results are averaged over 50 realizations, and the numbers in parentheses are standard deviations of the last digit. }
\end{table}

In applications, we are frequently faced with large-scale networks which exhibit more complicated topological characteristics than random graphs. Thus, it is more necessary and challenging to find a feasible strategy to efficiently approximate the optimal spreaders for those networks. Next, we explore CI-TM algorithm's performance for real networks. We examine two representative datasets: Youtube friendship network ($N=3,223,589, M=9,375,374, c=0.00138, l=5.29$) \cite{Mislove2009} and Internet autonomous system network ($N=1,464,020, M=10,863,640, c=0.00539, l=5.04$) \cite{Leskovec2005}. Here $N$ is the network size, $M$ is the number of undirected links, $c$ is the average clustering coefficient, and $l$ is the average shortest path length. Youtube network represents the undirected friend relations between users in the famous video sharing website Youtube. The Internet network records the communications between routers in different autonomous systems. The links between routers are constructed from the Border Gateway Protocol logs in an interval of 785 days. This provides an example of infrastructure network on which malicious attack and failure propagation may occur. Both networks are treated as undirected in the analysis.

\begin{figure}
\centering
\includegraphics[width=0.8\linewidth]{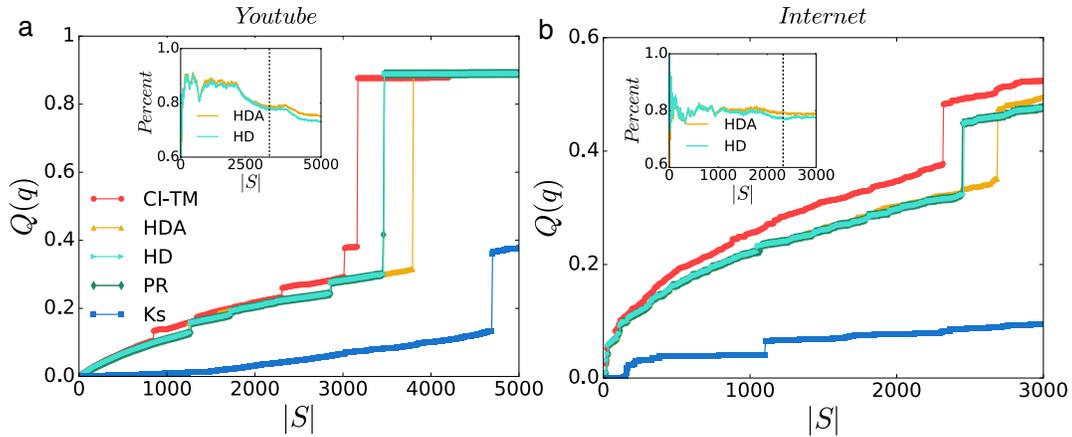}\\
\caption{{\bf Performance of CI-TM algorithm on large-scale real-world networks.} (a), The relationship between the size of active giant component $Q(q)$ and the number of seeds $|S|$ for Youtube friendship network, calculated by different methods. Inset displays the percentage of influencers predicted by CI-TM that HDA and HD have identified.The vertical dash line indicates the critical point of CI-TM. (b), Same analysis for Internet network.}\label{real}
\end{figure}

Figure \ref{real}a displays $Q(q)$ for different numbers of seeds $|S|$ for Youtube network. CI-TM is able to trigger the global cascade with a smaller group of seeds, whose size is quite small compared to the entire network due to the heavy-tailed degree distribution. We also discover that, in the setting of first-order transitions, some nodes with moderate numbers of connections play a crucial role in the collective influence of LTM. As shown in the inset figure, we present the percentage of influencers predicted by CI-TM that HDA and HD have identified, with the vertical dash line indicating CI-TM's $q_c$. At $q_c$, HDA and HD locate nearly $80\%$ overlapping seeds with CI-TM algorithm, most of which are tagged as hubs. However, due to the collective nature of LTM, seeding the set of privileged nodes in the non-interacting view does not guarantee the maximization of collective influence. The other proportion of spreaders with lower degree, although may be inefficient as single spreaders, are responsible for bridging the collective influence of hubs. With the help of both hubs and bridging low degree nodes, CI-TM can expand the collective influence with a smaller number of seeds. The Internet network also exhibits a similar phenomenon in Fig. \ref{real}b. In this case, HDA and HD can only find $80\%$ influencers at the first-order transition of CI-TM algorithm, missing a substantial amount of nodes with lower degree but indispensable in integrating the collective influence of high-degree seeds. 

Although the performance of K-core can be improved by adaptive calculation in Fig. \ref{random}, for large-scale real networks, we do not display the result of KsA due to its $O(N^2)$ computational complexity and only show the curve of K-core. One cause for the unsatisfactory result of K-core is that it is not designed as a multiple spreaders finder since high K-core nodes tend to form densely connected clusters in the same shell, which prevents the expanding of information cascade.

In Methods, we further compare CI-TM algorithm with other methods, including Betweenness Centrality (BC) \cite{Freeman1979}, Closeness Centrality (CC) \cite{Bavelas1950} and Greedy Algorithm (GA) \cite{Kempe2003}. Results from ER and SF networks suggest that CI-TM algorithm also outperforms computationally expensive BC, CC and GA.

\subsection*{Analysis of subcritical paths}

\begin{figure}
\centering
\includegraphics[width=0.8\linewidth]{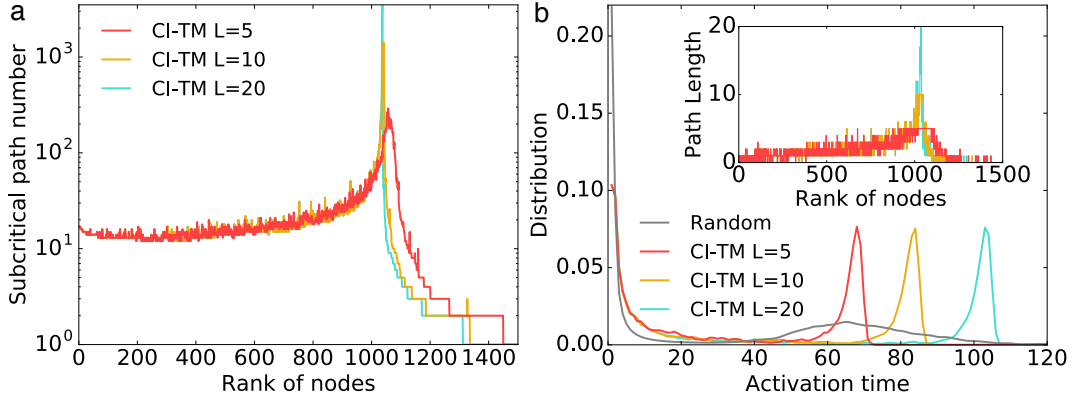}\\
\caption{{\bf Analysis of subcritical paths.} (a), Comparison of the number of subcritical paths attached to each node when it is activated sequentially according to CI-TM ranking. Results for CI-TM with $L=5$, $L=10$ and $L=20$ are displayed. LTM for $t=0.5$ runs on an ER network with size $N=10^4$ and average degree $\langle k\rangle=6$. (b), Distribution of activation time in the global cascading for CI-TM ($L=5, 10, 20$) and Random strategy at $q_c$. Inset shows the length of subcritical paths for each node when activated in CI-TM ranking. The curve for Random seed selection is averaged over 1,000 independent LTM realizations.}\label{LD}
\end{figure}

With the CI-TM algorithm, we present an analysis of subcritical paths in cascading process. In Fig. \ref{LD}a, we first display the evolution of the number of subcritical paths during the sequential activation process based on CI-TM ranking. We run LTM model for $t=0.5$ on an ER network with size $N=10^4$ and average degree $\langle k\rangle=6$. Nodes are activated sequentially according to their ranks in CI-TM algorithm. At the time of each activation, the number of subcritical paths attached to the node is calculated. After the activation, nodes on the subcritical paths are activated automatically, as we did in the CI-TM algorithm. In Fig. \ref{LD}a, the evolution of subcritical path number for CI-TM ($L=5, 10, 20$) is displayed. For all $L$ values, the number of subcritical paths peaks at the critical point, where the first-order transition occurs. In addition, as $L$ increases, the peak time of CI-TM is slightly shifted forward, while the peak value increases dramatically. As large numbers of subcritical paths imply a heavy computational burden, the majority of computation is concentrated around the critical point. Therefore, if we want to optimize the influence before the discontinuous transition, which is common in real-world applications, CI-TM algorithm becomes much more efficient since it avoids counting extremely long subcritical paths near the critical point.

We also examine the distribution of nodes' activation time in a global cascading. In Fig. \ref{LD}b, we report the distribution of activation time at the critical point for CI-TM ($L=5,10, 20$) and random selection of seeds. All the curves first decrease and then develop a second peak. Compared with the distribution of random selection, CI-TM has a much larger number of nodes getting activated at the second peak. More importantly, the increase of $L$ in CI-TM algorithm will postpone the arrival of the second peak, which is similar to the previous finding on regular networks \cite{Altarelli2013a}. In CI-TM algorithm with a larger $L$, longer subcritical paths are allowed during the calculation, as shown in the inset of Fig. \ref{LD}b. Considering the size of optimal seeds in Fig. \ref{LD}a and the distribution of activation time in Fig. \ref{LD}b, a smaller $L$ in CI-TM algorithm can expedite the global cascading, but at the expense of a few more seeds.

\section*{Discussion}

Identification of superspreaders in LTM has great practical implications in a wide range of dynamical processes. However, the complicated interactions among multiple spreaders prevent us from accurately locating the pivotal influencers in LTM. In this work, we propose a framework to analyze the collective influence of individuals in general LTM. By iteratively solving the linearized message passing equations, we decompose $\|\nu_{\to}\|$ into separate components, each of which corresponds to the contribution made by a single seed. Particularly, we find that the contribution of a seed is largely determined by its interplay with other nodes through subcritical paths.  In order to maximize the active population, we develop a scalable CI-TM algorithm that is feasible for large-scale networks. Results show that the proposed CI-TM algorithm outperforms other ranking strategies in synthetic random graphs and real-world networks. Our CI-TM algorithm can be employed in relevant applications such as viral marketing and information spreading in big-data analysis.

\section*{Methods}
\subsection*{Linearization of $G_{i\to j}$}
The conventional method to linearize the nonlinear function $G_{i\to j}(\nu_{\to})$ is Taylor expansion around the fixed point $\mathbf{0}$: $G_{i\to j}(\nu_{\to})\approx G_{i\to j}(\mathbf{0})+G'_{i\to j}(\mathbf{0})\nu_{\to}$. However, for our specific function $G_{i\to j}$, the gradient $G'_{i\to j}(\mathbf{0})$ is constantly $\mathbf{0}$ according to Eq. (\ref{calculateF}). To avoid the degeneracy, other linear approximation method should be applied.

For a differentiable function $f: \mathbb{R}^n\to\mathbb{R}$ and $\mathbf{x}$, $\mathbf{y}\in\mathbb{R}^n$, the mean value theorem guarantees that there exists a real number $c\in(0,1)$ such that $f(\mathbf{y})-f(\mathbf{x})=\nabla f((1-c)\mathbf{x}+c\mathbf{y})\cdot(\mathbf{y}-\mathbf{x})$. Here $\nabla$ is the gradient and $\cdot$ denotes the dot product. Set $f=G_{i\to j}$, $\mathbf{x}=\mathbf{0}$, and $\mathbf{y}=\nu_{\to}$, we have $G_{i\to j}(\nu_{\to})=G_{i\to j}(\mathbf{0})+G'_{i\to j}(c\nu_{\to})\nu_{\to}$. Notice that, if we set $c=0$, the above equation becomes the classical linear Taylor expansion: $G_{i\to j}(\nu_{\to})\approx G_{i\to j}(\mathbf{0})+G'_{i\to j}(\mathbf{0})\nu_{\to}$, where the approximation accuracy is $O(|\nu_{\to}|^2)$ ($|\cdot|$ is the norm of vectors). In a network with $N$ nodes and $M$ undirected links, we define the norm as $|\nu_{\to}|\equiv\sum_{ij}\nu_{i\to j}/2M$ ($2M$ is the number of directed links) so that $|\nu_{\to}|$ is bounded below 1 for network size $N\to\infty$.

To deal with the degeneracy of $G'_{i\to j}(\mathbf{0})$, we approximate $G_{i\to j}(\nu_{\to})$ by setting $c=1$ for small $\nu_{\to}$: $G_{i\to j}(\nu_{\to})\approx G_{i\to j}(\mathbf{0})+G'_{i\to j}(\nu_{\to})\nu_{\to}$. The approximation error can be calculated by $e=|G_{i\to j}(\nu_{\to})-G_{i\to j}(\mathbf{0})-G'_{i\to j}(\nu_{\to})\nu_{\to}|\leq|G'_{i\to j}(c\nu_{\to})-G'_{i\to j}(\nu_{\to})||\nu_{\to}|$. Recall that $G'_{i\to j}=(\cdots,\frac{\partial G_{i\to j}}{\partial\nu_{k\to\ell}}, \cdots)$. In a finite-size network, for a small $\nu_{\to}$ with elements $|\nu_{i\to j}|\leq1$, the gradient of each element $\frac{\partial G_{i\to j}}{\partial \nu_{k\to\ell}}$ is bounded according to Eq. (\ref{calculateF}). For all $2M$ elements, there exists a uniform upper bound for all the gradients $\nabla\frac{\partial G_{i\to j}}{\partial \nu_{k\to\ell}}$. Applying the mean value theorem to the differentiable function $\frac{\partial G_{i\to j}}{\partial \nu_{k\to\ell}}$, there should be a constant $L$ such that $|\frac{\partial G_{i\to j}}{\partial \nu_{k\to\ell}}(c\nu_{\to})-\frac{\partial G_{i\to j}}{\partial \nu_{k\to\ell}}(\nu_{\to})|\leq L|\nu_{\to}|$ for all the elements of  $G'_{i\to j}$. Summing up all the elements in $G'_{i\to j}$, we conclude that $|G'_{i\to j}(c\nu_{\to})-G'_{i\to j}(\nu_{\to})|\leq|(\cdots,L,\cdots)||\nu_{\to}|$. Therefore, the approximation error $e\leq|(\cdots,L,\cdots)||\nu_{\to}|^2$. This proves that the accuracy of the linear approximation is $O(|\nu_{\to}|^2)$, which is same as the linear Taylor expansion.

\begin{figure}
\centering
\includegraphics[width=0.5\columnwidth]{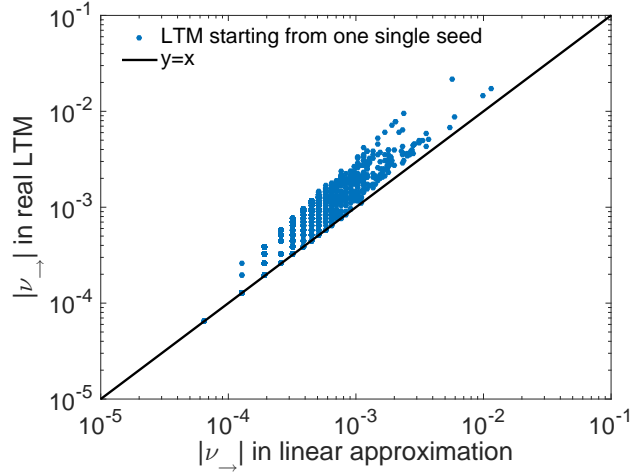}\\
\caption{{\bf Linear approximation of $|\nu_{\to}|$ for LTM initiated by one seed.} In a scale-free network ($N=5,000$, $M=7753$, $\gamma=3$), we run LTMs with threshold $t=0.5$ starting from each node. For each instance, the relationship between the real $|\nu_{\to}|$ value and its linear approximation is presented. }\label{approx}
\end{figure}

In the CI-TM algorithm, we only select one seed at each time step. Here we directly examine the accuracy of the linear approximation when LTM is initiated by one single seed. Specifically, we run LTM dynamics with threshold $t=0.5$ from each node in a scale-free network ($N=5,000$, $M=7753$, $\gamma=3$). For each of these realizations, we plot the realistic $|\nu_{\to}|$ value and its approximation in Fig. \ref{approx}. As expected, the approximation is generally lower than the real value since loops are neglected. The correlation between real values and approximations is 0.9118, and a higher correlation 0.9437 is obtained in the logarithmic scale. Therefore, the linear approximation in each step of CI-TM algorithm is accurate. As more seeds are considered, the approximation accuracy will decrease gradually. The decreasing rate should be related to the number of short loops existing in the network. How the density of short loops affects the approximation accuracy will be further explored in future works.

\subsection*{More comparisons with competing methods}

A growing number of methods aimed to rank nodes' influence in networks have been proposed in previous studies. Here we introduce some of the most widely used competing methods and perform a thorough comparison with CI-TM algorithm.

{\bf High degree (HD)} Degree, defined as the number of connections attached to a node, is the most widely-used measure of influence \cite{Albert2000}. In HD method, we rank nodes according to their degrees in a descending order, and sequentially select them as information sources. For HD method, the selected hubs intend to link with each other due to the assortative mixing property, making their influence areas overlap significantly. In this case, the selected seeds could rarely be optimal. High degree adaptive {\bf (HDA)} is the adaptive version of HD method. After each removal, the degree of each node is recalculated. Such adaptive procedure can usually mitigate the overlapping and improve the performance of HD.

{\bf K-core (Ks)} Through a k-shell decomposition process, K-core method assigns each node a $k_S$ value to distinguish whether it locates in the core region or peripheral area \cite{Seidman1983}. In k-shell decomposition, nodes are iteratively removed from the network according to their current degrees. During the removal, all the nodes are classified into different k-shells. The K-core method selects nodes within high k-shells as the spreaders. In practice, single influential spreaders can be identified effectively by K-core ranking, which has been confirmed by both simulations and real-world data \cite{Kitsak2010,Pei2014,Pei2013}. However, K-core ranking has the disadvantage of severe overlap of seeds' influence areas, and therefore performs poorly for multiple node selection. This limitation can be alleviated with an adaptive scheme where we recompute the K-core after each removal. Since there exists many nodes in the same k-shell, we select the node with the largest degree to further distinguish nodes within the highest k-shell. Such K-core adaptive {\bf (KsA)} method can effectively enhance the performance of K-core.

{\bf PageRank (PR)} PageRank is a popular ranking algorithm of webpages which was developed and used by search engine Google \cite{Brin1998}. Over the years, PageRank has been adopted in many practical ranking problems. Generally speaking, PageRank measures a webpage's stationary visiting probability by a random walker following the hyperlinks in the network. As a special case of eigenvector centralities, PageRank evaluates the score of a node by taking into account its neighbors' scores. Even though such score-propagating mechanism works well for some purposes such as webpage ranking, an unfavorable consequence may be a heavy accumulation of scores near the high-degree nodes, specially for scale-free networks \cite{Martin2014}.

{\bf Greedy algorithm (GA)} In GA, starting from an empty set of seeds, nodes with the maximal marginal gain are sequentially added to the seed set. Kempe {\it et al.} have proven that for a class of LTM with the attribute of submodularity, GA has a performance guarantee of $1-1/e\approx 63\%$, which means it could achieve at least $63\%$ of real maximal influence \cite{Kempe2003}. This result relies on the submodular property defined by a diminishing returns effect: the marginal gain from adding a node to the seed set $S$ decreases with the size of $S$. It has been proven that several classes of LTM have submodular property, such as  a random choice of thresholds. However, for LTM with a fixed threshold, it is not generally submodular. As a consequence, GA is not guaranteed to provide a such approximation of the optimal spreaders for general LTM. Furthermore, the greedy search of GA requires massive simulations, which makes GA unscalable and thus limits its application in large-scale social networks.

{\bf Betweenness centrality (BC)} BC quantifies the importance of node $i$ in terms of the number of shortest paths cross through it \cite{Freeman1979}. Therefore, nodes with large BC usually occupy the pivotal positions in the shortest pathways connecting large numbers of nodes. In BC method, we select nodes with high BC scores as seeds. Although BC has been widely applied in social network analysis, its relatively high computational complexity makes BC prohibitive for large-scale networks. A typical BC algorithm takes $O(MN)$ to calculate for a network with $N$ nodes and $M$ links \cite{Brandes2001}, which is still not applicable to modern social networks with millions of users.

{\bf Closeness centrality (CC)} Closeness centrality quantifies how close a node to other nodes in the network \cite{Bavelas1950}. Formally, CC is defined as the reciprocal of the average shortest distance of a node to others in a network. Thus, nodes with high CC values tend to locate at the center of network clusters or communities. In CC method, we pick the seeds according to nodes' CC scores. Same as BC, CC also requires the heavy task of calculating all possible shortest paths. Thus the high computational cost of CC makes it infeasible for large-scale networks.

{\bf Max-Sum (MS)} F. Altarelli {\it et al.} developed a Max-Sum algorithm aimed to find the initial configurations that maximized the final number of active nodes in threshold models. Precisely, the trajectory of nodes' states is parametrized by a configuration $\mathbf{t}=\{t_i, 1\leq i\leq N\}$ where $t_i\in\mathcal{T}=\{0, 1,\cdots, T, \infty\}$ is the activation time of node $i$ ($t_i=\infty$ if inactive). By mapping the optimization onto a constraint satisfaction problem, an energy-minimizing algorithm based on the cavity method of statistical physics is proposed. In the algorithm, a convolution process is employed to compute the Max-Sum updates. The technical details of the derivation and implementation of the MS algorithm can be found in Ref. \cite{Altarelli2013}. In some cases MS algorithm does not converge, then a reinforcement strategy is implemented \cite{Altarelli2013}. By imposing an external field slowly increasing over time with a growth rate $r$, the system is forced to converge to a higher energy, which increases with $r$. In addition, it requires $O(r^{-1})$ iterations to reach convergence. For a node of degree $k$ and threshold $m_i$, each update takes $O(Tk(k-1)m_i^2)$ operations \cite{Altarelli2013}. Pre-computing the convolution can further save a factor of $k-1$. Considering the updates of all $N$ nodes for $O(r^{-1})$ iterations, the overall complexity of MS is $O(T\sum_i k_im_i^2/r)$. Therefore, the time complexity of MS depends on both the degree distribution of networks and the choice of threshold.

\begin{figure}
\centering
\includegraphics[width=0.8\columnwidth]{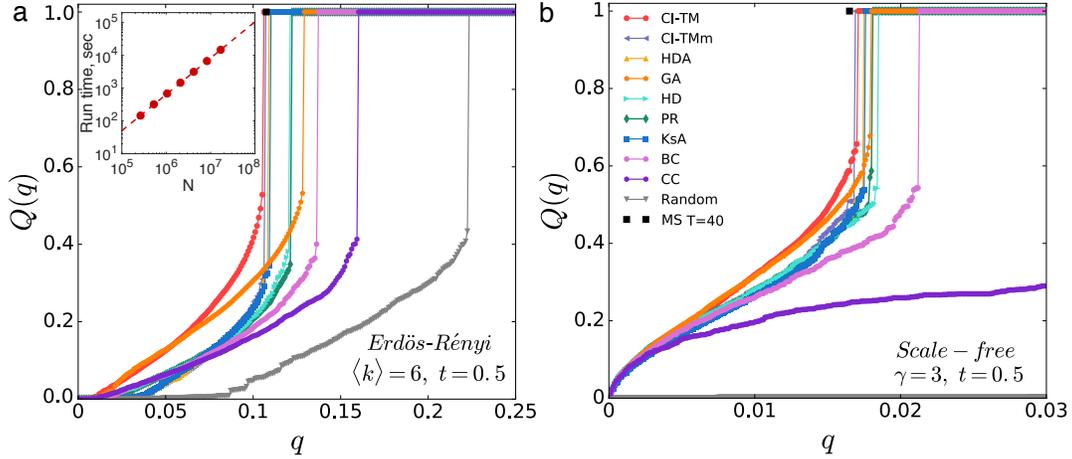}\\
\caption{{\bf Comparison of competing methods.} Performance of different methods for (a) ER network ($N=10^4$, $\langle k\rangle=6$, $t=0.5$) and  (b) scale-free network ($N=10^4$, $\gamma=3$, $t=0.5$). In addition to the methods we have examined in the main text, we also display the results of Greedy Algorithm (GA), Betweenness Centrality (BC), Closeness Centrality (CC) and Max-Sum (MS). We set the parameter $T=40$ in MS and implement reinforcement with parameter $r=10^{-4}$. The inset displays the run time of CI-TM ($L=3$) on ER networks ($\langle k\rangle=6$, $t=0.5$). The dashed line is power-law fitting.}\label{mp}
\end{figure}

In Fig. \ref{mp}, we provide the thorough comparisons of different methods on ER and SF networks ($N=10^4$), including computationally expensive methods GA, BC and CC. We set $T=40$ and a reinforcement parameter $r=1\times10^{-4}$ in MS algorithm. For both homogeneous and heterogeneous networks, CI-TM shows a larger active population for a given fraction of seed $q$ compared with other heuristic ranking strategies. We display the scaling of run time of CI-TM algorithm for ER networks with $\langle k\rangle=6$ and threshold $t=0.5$ as a function of $N$ in the inset of Fig. \ref{mp}a. CI-TM algorithm with $L=3$ is feasible for networks with size up to $N=O(10^8)$ within run time of $O(10^5)$ seconds, which can be applied to the modern large-scale online social networks. 

\subsection*{Modified CI-TM algorithm}

\begin{figure}
\centering
\includegraphics[width=0.5\columnwidth]{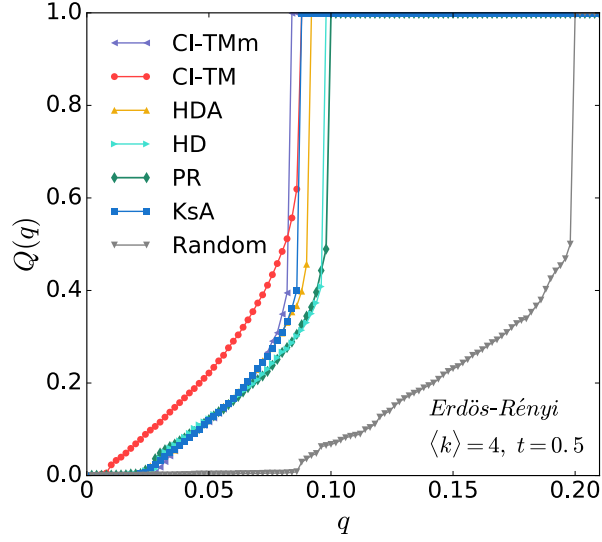}\\
\caption{{\bf Performance of modified CI-TM on ER networks.} For ER networks ($N=2\times10^5$, $\langle k\rangle=4$, $t=0.5$), the performance of CI-TMm surpasses CI-TM by excluding the fragmented vulnerable clusters in the adaptive calculation. Although $Q(q)$ for CI-TMm is lower at first, it exceeds other methods near the critical point and achieves the earliest first-order transition.}\label{er4}
\end{figure}

The CI-TM algorithm is essentially a greedy approach based on CI-TM values. The success of CI-TM algorithm depends on whether the currently selected seed has potentials to create more subcritical nodes that are helpful for the early formation of giant subcritical cluster. In LTM, there exists a special case of subcritical nodes with threshold $m=1$, which is defined as vulnerable vertices in previous literature \cite{Watts2002}. Different from general subcritical nodes, vulnerable vertices are naturally subcritical since they have threshold $m=1$ and do not rely on the states of others. During the calculation, a node of degree $k$ becomes vulnerable when its $m-1$ neighbors are removed, leaving $k-m+1$ links in the network. For ER networks with a low average degree, the limited number of remaining links of vulnerable vertices could only form fragmented clusters. In this case, CI-TM would bias to nodes connected to large numbers of small vulnerable clusters, such as a peripheral hub linked to numerous leaf nodes. Unfortunately, the activation of such small clusters provides little help to the formation of giant subcritical cluster. Because the scattered vulnerable clusters have very few links connected to existing non-subcritical nodes, their activations are not effective in producing subsequent subcritical nodes. Moreover, once global cascade appears, these fragmented vulnerable clusters will be activated subsequently, without additional seeds. In this case, we heuristically propose a modified CI-TM (CI-TMm) algorithm by excluding vulnerable nodes in the calculation of CI-TM value. The performance of CI-TMm algorithm is displayed in Fig. \ref{er4} for ER networks with an average degree $\langle k\rangle=4$. The critical value $q_c$ for CI-TMm is advanced compared to CI-TM algorithm. However, before the first-order transition, CI-TMm cannot optimize the spreading and has substantially lower $Q(q)$ than CI-TM. We should note that, CI-TMm presents a lower $q_c$ only in the situation of fragmented vulnerable clusters. For ER networks with higher average degrees (e.g., $\langle k\rangle=6$) where relatively large vulnerable clusters emerge, CI-TM still predicts earlier first-order transition.

\section*{Acknowledgements}
This work was supported by NIH-NIBIB 1R01EB022720, NIH-NCI  U54CA137788/ U54CA132378, NSF-PoLS PHY-1305476, NSF-IIS 1515022, and ARL Cooperative Agreement Number W911NF-09-2-0053, the ARL Network Science CTA (to H.A.M.), as well as US NIH grant GM110748 and the Defense Threat Reduction Agency contract HDTRA1-15-C-0018 (to J.S.).

\section*{Author contributions statement}
S.P., X.T, J.S., F.M. and H.A.M. designed research, performed study, analyzed data, and wrote the paper. All authors reviewed the manuscript.

\section*{Additional information}
\subsection*{Competing financial interests}
The authors declare no competing financial interests.


\begin{thebibliography}{9}

\bibitem{Buldyrev2010}
Buldyrev, S. V., Parshani, R., Paul, G., Stanley, H. E. \& Havlin, S. Catastrophic cascade of failures in interdependent networks. {\it Nature} {\bf 464,} 1025-1028 (2010).

\bibitem{Watts2007a}
Watts, D. J. \& Dodds, P. S. Influentials, networks, and public opinion formation. {\it J. Consum. Res.} {\bf 34,} 441-458 (2007).

\bibitem{Rogers1995}
Rogers, E. M. {\it Diffusion of Innovation} (Free Press, New York, 1995).

\bibitem{Reis2014}
Reis, S. D. {\it et al.} Avoiding catastrophic failure in correlated networks of networks. {\it Nature Phys.} {\bf 10,} 762-767 (2014) .

\bibitem{Kleinberg2007}
Kleinberg, J. {\it Algorithmic Game Theory (Cascading Behavior in Networks: Algorithmic and Economic Issues)} (Cambridge University Press, Cambridge, 2007) chapter 24, 613-632.

\bibitem{Domingos2001}
Domingos, P \& Richardson, M. Mining the network value of customers. In {\it Proc. 7th ACM SIGKDD Int. Conf. on Knowledge Discovery and Data Mining}, 57-66 (ACM, 2001).

\bibitem{Valente1999}
Valente, T. W. \& Davis, R. L. Accelerating the diffusion of innovations using opinion leaders. {\it Ann. Am. Acad. Polit. Soc. Sci.} {\bf 556,} 55-67 (1999).

\bibitem{Galeotti2009}
Galeotti, A. \& Goyal, S. Influencing the influencers: a theory of strategic diffusion. {\it The RAND J. Econ.} {\bf 40,} 509-532 (2009).


\bibitem{Kempe2003}
Kempe, D., Kleinberg, J. \& Tardos, \'{E}. Maximizing the spread of influence through a social network. In {\it Proc. 9th ACM SIGKDD Int. Conf. on Knowledge Discovery and Data Mining}, 137-146 (ACM, 2003).

\bibitem{Leskovec2007}
Leskovec, J., Krause, A., Guestrin, C., Faloutsos, C., VanBriesen, J. \& Glance, N. Cost-effective outbreak detection in networks. In {\it Proc. 13th ACM SIGKDD Int. Conf. on Knowledge Discovery and Data Mining}, 420-429 (ACM, 2007).

\bibitem{Chen2009}
Chen, W., Wang, Y. \& Yang, S. Efficient influence maximization in social networks. In {\it Proc. 15th ACM SIGKDD Int. Conf. on Knowledge Discovery and Data Mining}, 199-208 (ACM, 2009).


\bibitem{Kitsak2010}
Kitsak, M. {\it et al.} Identification of influential spreaders in complex networks. {\it Nature Phys.} {\bf 6,} 888-893 (2010).

\bibitem{Pei2013}
Pei, S. \& Makse, H. A. Spreading dynamics in complex networks. {\it J. Stat. Mech.} {\bf 12,} P12002 (2013).

\bibitem{Pei2014}
Pei, S., Muchnik, L., Andrade Jr, J. S., Zheng, Z. \& Makse, H. A. Searching for superspreaders of information in real-world social media. {\it Sci. Rep.} {\bf 4,} 5547 (2014).

\bibitem{Morone2015}
Morone, F. \& Makse, H. A. Influence maximization in complex networks through optimal percolation. {\it Nature} {\bf 524,} 65-68 (2015).

\bibitem{Morone2016}
Morone, F., Min, B., Bo, L., Mari, R. \& Makse, H. A. Collective Influence Algorithm to find influencers via optimal percolation in massively large social media. {\it Sci. Rep.} {\bf 6,} 30062 (2016).
\bibitem{Altarelli2013a}
Altarelli, F., Braunstein, A., Dall'Asta, L. \& Zecchina, R. Large deviations of cascade processes on graphs. {\it Phys. Rev. E} {\bf 87,} 062115 (2013).
\bibitem{Altarelli2013}
Altarelli, F., Braunstein, A., Dall'Asta, L. \& Zecchina, R. Optimizing spread dynamics on graphs by message passing. {\it J. Stat. Mech.} {\bf 9,} P09011  (2013).

\bibitem{Guggiola2015}
Guggiola, A. \& Semerjian, G. Minimal contagious sets in random regular graphs. {\it J. Stat. Phys.} {\bf 158,} 300-358 (2015).

\bibitem{Mugisha2016}
Mugisha, S. \& Zhou, H. J. Identifying optimal targets of network attack by belief propagation. {\it Phys. Rev. E} {\bf 94,} 012305 (2016).

\bibitem{Braunstein2016}
Braunstein, A., Dall'Asta, L., Semerjian, G. \& Zdeborov\'{a}, L. Network dismantling. {\it Proc. Natl. Acad. Sci. USA} {\bf 113,} 12368-12373 (2016).

\bibitem{Teng2016}
Teng, X., Pei, S., Morone, F. \& Makse, H. A. Collective influence of multiple spreaders evaluated by tracing real information flow in large-scale social networks. {\it Sci. Rep.} {\bf 6,} 36043 (2016).
\bibitem{Pei2015}
Pei, S., Muchnik, L., Tang, S., Zheng, Z. \& Makse, H. A. Exploring the complex pattern of information spreading in online blog communities. {\it PLoS ONE} {\bf 10,} e0126894 (2015).

\bibitem{Radicchi2016}
Radicchi, F. \& Castellano, C. Leveraging percolation theory to single out influential spreaders in networks. {\it Phys. Rev. E} {\bf 93,} 062314 (2016).
\bibitem{Hu2015}
Hu, Y., Ji, S., Feng, L. \& Jin, Y. Quantify and maximise global viral influence through local network information. {\it arXiv preprint} arXiv:1509.03484 (2015).

\bibitem{Lawyer2015}
Lawyer, G. Understanding the influence of all nodes in a network. {\it Sci. Rep.} {\bf 5,} 8665 (2015).

\bibitem{Quax2013}
Quax, R., Apolloni, A. \& Sloot, P. M. The diminishing role of hubs in dynamical processes on complex networks. {\it J. R. Soc. Interface} {\bf 10,} 20130568 (2013).

\bibitem{Tang2015}
Tang, S., Teng, X., Pei, S., Yan, S. \& Zheng, Z. Identification of highly susceptible individuals in complex networks. {\it Physica A} {\bf 432,} 363-372 (2015).

\bibitem{Albert2000}
Albert, R., Jeong, H. \& Barab\'{a}si, A. L. Error and attack tolerance of complex networks. {\it Nature} {\bf 406,} 378-382 (2000).


\bibitem{Freeman1979}
Freeman, L. C. Centrality in social networks conceptual clarification. {\it Soc. Netw.} {\bf 1,} 215-239 (1978).

\bibitem{Brin1998}
Brin, S. \& Page, L. Reprint of: The anatomy of a large-scale hypertextual web search engine. {\it Computer networks} {\bf 56,} 3825-3833 (2012).

\bibitem{Seidman1983}
Seidman, S. B. Network structure and minimum degree. {\it Soc. Netw.} {\bf 5,} 269-287 (1983).

\bibitem{Granovetter1978}
Granovetter, M. Threshold models of collective behavior. {\it Am. J. Sociol.} {\bf 83,} 1420-1443 (1978).

\bibitem{Schelling1978}
Schelling, T. C. {\it Micromotives and macrobehavior} (Norton, New York, 1978).
\bibitem{Valente1995}
Valente, T. W. {\it Network Models of the Diffusion of Innovations} (Hampton Press, Cresskill, NJ, 1995).
\bibitem{Watts2002}
Watts, D. J. A simple model of global cascades on random networks. {\it Proc. Natl. Acad. Sci. USA} {\bf 99,} 5766-5771 (2002).

\bibitem{Baxter2010}
Baxter, G. J., Dorogovtsev, S. N., Goltsev, A. V. \& Mendes, J. F. F. Bootstrap percolation on complex networks. {\it Phys. Rev. E} {\bf 82,} 011103 (2010).

\bibitem{Goltsev2006}
Goltsev, A. V., Dorogovtsev, S. N. \& Mendes, J. F. F. k-core (bootstrap) percolation on complex networks: Critical phenomena and nonlocal effects. {\it Phys. Rev. E} {\bf 73,} 056101 (2006).
\bibitem{Dorogovtsev2006}
Dorogovtsev, S. N., Goltsev, A. V. \& Mendes, J. F. F. k-core architecture and k-core percolation on complex networks. {\it Physica D} {\bf 224,} 7-19 (2006).
\bibitem{Schwarz2006}
Schwarz, J. M., Liu, A. J. \& Chayes, L. Q. The onset of jamming as the sudden emergence of an infinite k-core cluster. {\it Europhys. Lett.} {\bf 73,} 560 (2006).

\bibitem{Melnik2011}
Melnik, S., Hackett, A., Porter, M. A., Mucha, P. J. \& Gleeson, J. P. The unreasonable effectiveness of tree-based theory for networks with clustering. Phys. Rev. E, {\bf 83,} 036112  (2011).

\bibitem{Hashimoto1989}
Hashimoto, K. Zeta functions of finite graphs and representations of p-adic groups. {\it Adv. Stud. Pure Math.} {\bf 15,} 211 (1989).
\bibitem{Martin2014}
Martin, T., Zhang, X. \& Newman, M.E.J. Localization and centrality in networks. {\it Phys. Rev. E} {\bf 90,} 052808 (2014).


\bibitem{Molloy1995}
Molloy, M. \& Reed, B. A critical point for random graphs with a given degree sequence. {\it Random Structures \& Algorithms} {\bf 6,} 161-180 (1995).

\bibitem{Mislove2009}
Mislove, A. Online Social Networks: Measurement, Analysis, and Applications to Distributed Information Systems. {\it PhD thesis, Rice University} (2009).

\bibitem{Leskovec2005}
Leskovec, J., Kleinberg, J. \& Faloutsos, C. Graphs over time: densification laws, shrinking diameters and possible explanations. In {\it Proc. 11th ACM SIGKDD Int. Conf. on Knowledge Discovery and Data Mining}. 177-187 (ACM, 2005).

\bibitem{Bavelas1950}
Bavelas, A. Communication patterns in tasks oriented groups. {\it J. Acoust. Soc. Am.} {\bf 22,} 725-730 (1950).

\bibitem{Brandes2001}
Brandes, U. A faster algorithm for betweenness centrality. {\it J. Math. Sociol.} {\it 25,} 163-177 (2001).

\end{thebibliography}
\end{document}